\documentclass[draft]{agujournal2018}
\usepackage{apacite}
\usepackage{url} 
\usepackage{lineno}
\usepackage{gensymb}
\journalname{JGR: Space Physics}

\begin{document}


\title{ Magnetohydrodynamic with embedded particle-in-cell simulation of the Geospace Environment Modeling dayside kinetic processes challenge event }
\authors{Yuxi Chen\affil{1},
  {G\'abor T\'oth} \affil{1},
  Heli Hietala\affil{2,3,4},
  Sarah K. Vines\affil{5},
  Ying Zou\affil{6},
  Yukitoshi Nishimura\affil{7},
  Marcos V.D. Silveira\affil{8,9},
  Zhifang Guo\affil{10},
  Yu Lin\affil{10},
  Stefano Markidis\affil{11},
  }

\affiliation{1}{Department of Climate and Space Sciences and Engineering, University of Michigan, , Ann Arbor, MI, USA}
\affiliation{2}{Space Research Laboratory, Department of Physics and Astronomy, University of Turku, Finland}
\affiliation{3}{Department of Earth, Planetary and Space Sciences, University of California, Los Angeles, CA, USA}
\affiliation{4}{The Blackett Laboratory, Imperial College London, London, UK}
\affiliation{5}{Johns Hopkins University Applied Physics Laboratory, Laurel, MD, USA}
\affiliation{6}{The Center for Space Plasma and Aeronomic Research, University of Alabama, Huntsville, AL, USA}
\affiliation{7}{Department of Electrical and Computer Engineering and Center for Space Physics, Boston University, Boston, MA, USA}
\affiliation{8}{NASA Goddard Space Flight Center, Greenbelt, MD, USA}
\affiliation{9}{Catholic University of America, Washington DC, USA}
\affiliation{10}{Physics Department, Auburn University, Auburn, AL, USA}
\affiliation{11}{KTH, Stockholm, Sweden}

\correspondingauthor{Yuxi Chen}{yuxichen@umich.edu}

\begin{keypoints}
  \item 1 The MHD-EPIC simulation magnetic fields and plasma data match MMS3 observations well during the magnetopause crossing
  \item 2 There are usually multiple X-lines at the magnetopause in the MHD-EPIC simulation
  \item 3 The MHD-EPIC simulation shows complex movement and spreading of the X-lines
\end{keypoints}

\begin{abstract}
  We use the MHD with embedded particle-in-cell model (MHD-EPIC) to study the Geospace Environment Modeling (GEM) dayside kinetic processes challenge event at 01:50-03:00 UT on 2015-11-18, when the magnetosphere was driven by a steady southward IMF. In the MHD-EPIC simulation, the dayside magnetopause is covered by a PIC code so that the dayside reconnection is properly handled. We compare the magnetic fields and the plasma profiles of the magnetopause crossing with the MMS3 spacecraft observations. Most variables match the observations well in the magnetosphere, in the magnetosheath, and also during the current sheet crossing. The MHD-EPIC simulation produces flux ropes, and we demonstrate that some magnetic field and plasma features observed by the MMS3 spacecraft can be reproduced by a flux rope crossing event. We use an algorithm to automatically identify the reconnection sites from the simulation results. It turns out that there are usually multiple X-lines at the magnetopause. By tracing the locations of the X-lines, we find the typical moving speed of the X-line endpoints is about 70~km/s, which is higher than but still comparable with the ground-based observations.  
\end{abstract}

\section{Introduction}
The dayside magnetopause reconnection is the most important mechanism for the mass and energy transfer from the solar wind to Earth's magnetosphere. Since the magnetic field in the magnetosphere is usually stronger than the magnetosheath magnetic field, the dayside reconnection is asymmetric. The processes of the dayside asymmetric reconnection have been studied with both spacecraft data and numerical models. 

Particle-in-cell (PIC) codes have been widely used to investigate the kinetic proprieties of the asymmetric reconnection, such as the reconnection rate \citep{Cassak:2007}, the electric field and magnetic field structures \citep{Mozer:2008, Malakit:2013}, and the signatures of the electron diffusion regions \citep{Shay:2016}. On the other hand, the efficient MHD models are well-suited for investigating the global features of the magnetopause reconnection. For example, \citet{Borovsky:2008} studied the global reconnection rate with the global MHD model BATS-R-US \citep{Powell:1999}, and \citet{Komar:2014} compared the global MHD simulations with several dayside magnetic reconnection location models \citep{Moore:2002, Trattner:2007}. In recent years, more and more kinetic models are applied to simulate the kinetic processes at the magnetopause, such as the hybrid models \citep{Tan:2011, Karimabadi:2014}, the hybrid-Vlasov model \citep{Hoilijoki:2017}, and the MHD with embedded particle-in-cell (MHD-EPIC) model \citep{Chen:2017}.  

As the products of the dayside magnetopause reconnection, the flux transfer events (FTEs) have attracted the attention of the numerical modeling community. Ideal-MHD \citep{Fedder:2002, Raeder:2006,Sibeck:2008} and resistive MHD \citep{Dorelli:2009} models have been used to generate FTEs in global simulations. Recently, more sophisticated models that contain kinetic physics have also been used to study the FTEs. \citet{Hoilijoki:2017} performed a 2D global magnetospheric hybrid-Vlasov simulation to investigate the dayside reconnection and FTEs. \citet{Chen:2017} studied the generation and evolution of FTEs with 3D MHD-EPIC model. 

Another prominent topics of the 3D dayside reconnection is the spreading of the X-lines. \citet{Huba:2002} found the X-line in a Hall-MHD simulation propagates asymmetrically along the current channel like a wave. The growth of the X-line was further studied by a hybrid code \citep{Karimabadi:2004} and a two fluid code \citep{Shay:2003}. From 3D PIC simulations, \citet{Lapenta:2006} found the X-line grows in the direction of the current carrier, and the X-line spreading speed depends on the current sheet thickness. \citet{Shepherd:2012} discussed the role of the guide field. They suggested the X-line spreading is due to the motion of the current carrier under weak guide field, and the bidirectional spreading is caused by the Alfven waves along the guide field. Recently, the X-line spreading at the magnetopause is observed by the SuperDARN radar \citep{Zou:2018}. The SuperDARN observations suggested the X-line spreading speed is about 40~km/s for the reconnection under weak guide field. 

Numerical simulations are crucial for understanding the dynamics at the magnetopause. To assess the performance of the numerical models on the dayside kinetic processes, the Geospace Environment Modeling (GEM) dayside kinetic processes focus group combined efforts from both modelers and observers to study the same event. The focus group selected the southward IMF event on 2015-11-18 01:50-03:00 UT as the challenge event. This challenge is a collaborative effort by both numerical modelers and observers to compare the numerical simulation results with the spacecraft and ground-based observations. \citet{Kitamura:2016} has analyzed the MMS and Geotail data for this event, and estimated the X-line location to be around $Z_{GSM} = 2~R_E$. Recently, \citet{Nishimura:2020} studied the X-line spreading of this event. 
We use the MHD-EPIC model \citep{Daldorff:2014} to simulate the challenge event in the present paper. Compared to the study by \citet{Chen:2017}, the present paper uses a realistic dipole field and solar wind conditions so that the simulation results are comparable to the observations, and a new robust and accurate particle-in-cell algorithm \citep{Chen:2019} is used to improve the simulation quality. In this paper, we focus on the model-data comparisons. We compare the magnetopause crossing magnetic field and plasma data with the MMS3 data, and show the movement and spreading of the X-lines in the simulation are comparable to the ground-based observations. 

In the following section, the numerical details of the MHD-EPIC model are described, and section~\ref{section:res} presents the simulation results and compares the simulation with observations. 

\section{Numerical models}
\label{section:model}
The MHD-EPIC model \citep{Daldorff:2014}, which two-way couples the Hall-MHD model BATS-R-US \citep{Powell:1999, Toth:2008} and the semi-implicit particle-in-cell code iPIC3D \citep{Markidis:2010,Chen:2019} through the Space Weather Modeling Framework (SWMF) \citep{Toth:2005swmf, Toth:2012swmf}, is applied to study the challenge event on 2015-11-18. The dayside magnetopause is covered by the particle-in-cell (PIC) code so that the kinetic effects of the dayside magnetic reconnection are incorporated into the model, and the fluid model BATS-R-US handles the rest of the simulation domain. The MHD-EPIC simulation in the present paper uses the same fluid model, i.e., the Hall-MHD model with a separate electron pressure equation, and the same boundary condition types as the simulation performed by \citet{Chen:2017}. But the dipole field, the inner boundary density, and the solar wind conditions are different from those of \citet{Chen:2017}. The dipole field is approximately $27^{\circ}$ tilted from the $Z_{GSM}$-axis towards the negative $X_{GSM}$-direction. The present paper uses a fixed inner boundary density of 8~amu/cc at $r=2.5~R_E$ to match the magnetospheric plasma profiles that were observed by the MMS satellites (Figure~\ref{fig:plot2}). A steady solar wind with $\mathbf{B}=(0,0,-6)$ nT, mass density $\rho = 9.5~\mathrm{amu/cm^3}$, ion temperature $T_i=9~\mathrm{eV}$, electron temperature $T_e=9~\mathrm{eV}$, and solar wind velocity $\mathbf{u}=(-365,0,0)$ km/s, is used to drive the magnetosphere. These solar wind values are obtained by averaging and simplifying the ACE and Wind satellites data. In this simulation, BATS-R-US uses a locally refined Cartesian grid with a cell size of $1/16~R_E$ around the dayside magnetopause.

\begin{figure}
  \includegraphics[width=0.9\textwidth, trim=0cm 0cm 0cm 0cm,clip,angle=0]{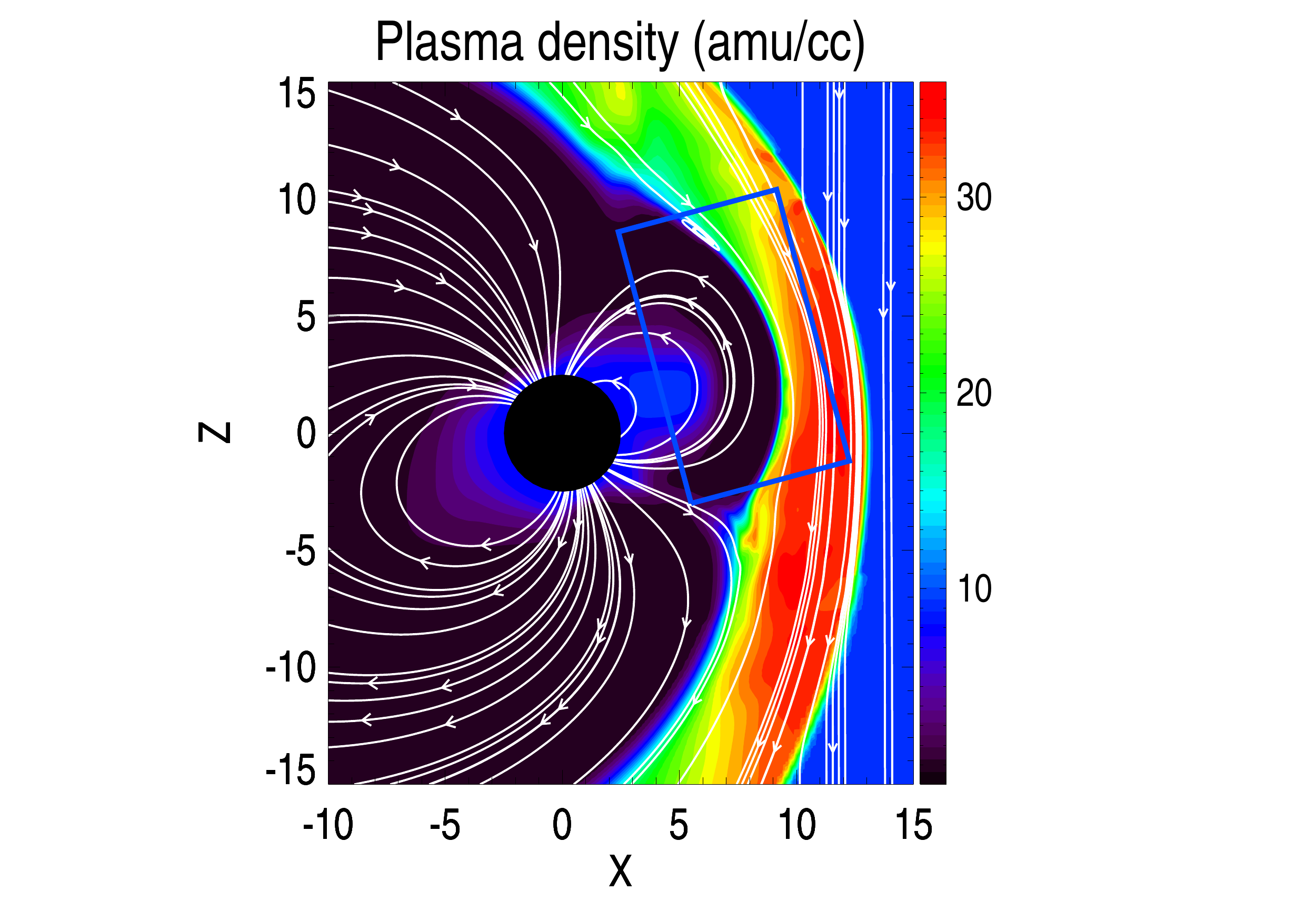}
  \caption{The plasma density and the magnetic field lines in the $Y_{GSM}=0$ plane. 
  The blue rectangular box represents the region that is simulated by the PIC code.}
  \label{fig:pic_box}
\end{figure}

The PIC code uses the latest Gauss’s Law satisfying Energy Conserving Semi-Implicit Method (GL-ECSIM) \citep{Chen:2019}, and it covers the dayside magnetopause (Figure~\ref{fig:pic_box}). The PIC region is rotated $15^{\circ}$ from the $Z_{GSM}$-axis to the $X_{GSM}$-axis to be aligned with the dayside magnetopause. The size of the PIC box is $L_x=7~R_E$, $L_y=16~R_E$ and $L_z=12~R_E$. It extents from $-8~R_E$ to $8~R_E$ in the GSM-Y direction. In the GSM X-Z plane, its bottom-left corner is at $x=5.5~R_E$ and $z=-3~R_E$, and the rotation is performed around this corner. After the rotation, the Y-axis of the PIC coordinates is still parallel with $Y_{GSM}$, but the X-axis and the Z-axis of the PIC domain are not aligned with the GSM coordinates anymore. The transformation between the PIC coordinates and the GSM coordinates in the units of $R_E$ are:

\begin{eqnarray}
  \label{eq:coord}
X_{GSM} &&= X_{PIC}\cdot\cos(15^{\circ}) - Z_{PIC} \cdot \sin(15^{\circ}) + 5.5\\
Y_{GSM} &&= Y_{PIC} - 8\\
Z_{GSM} &&= X_{PIC}\cdot\sin(15^{\circ}) + Z_{PIC} \cdot \cos(15^{\circ}) - 3.
\end{eqnarray}

A uniform Cartesian mesh with a cell size of $1/25~R_E$ is used for the PIC simulation. 100 macro-particles per species per cell are applied as the initial conditions and the boundary conditions. The physical ion inertial length $d_i$ is just about $40\,\mathrm{km/s}$ in the magnetosheath, and it is extremely expensive to resolve such a small scale in a global simulation. So, similar to the simulation by \citet{Chen:2017}, we artificially increase the plasma kinetic scales by a factor of 16 by reducing the charge per mass ratio \citep{Toth:2017}. The electron kinetic scales are further increased by using a reduced ion-electron mass ratio of $m_i/m_e=100$. In the magnetosheath, the mesh resolves one inertial length with about three cells, which is coarser than typical PIC simulations due to the limitation of the computational resources. The grid resolution is not high enough to well resolve the electron scales, e.g. electron skin depth, and some kinetic processes related to magnetic reconnection, such as the particle-wave interaction and streaming instability, may not be described accurately. In the following section, we show that the MHD-EPIC simulation still agrees with MMS observations well in general. We focus on the MHD-EPIC simulation results in this paper, but we also present the ideal-MHD and Hall-MHD simulations for comparison. We run the model BATS-R-US with the ideal-MHD equations first with the local time-stepping scheme to reach a steady-state, and then continue with a 1-hour simulation in time-accurate mode to make the magnetopause structures sharper. This ideal-MHD simulation results at $t=1~\mathrm{h}$ is used as the initial conditions of the 3-hour-long (from $t=1~\mathrm{h}$ to $t=4~\mathrm{h}$) MHD-EPIC and Hall-MHD simulations. Ideal-MHD itself also runs to $t=4~\mathrm{h}$ for comparison. We use the simulation results from $t=1~\mathrm{h}$ to $t=4~\mathrm{h}$ for the analyses in the next section. In the pure Hall-MHD simulation, the ion inertial length is also artificially increased by a factor of 16 by reducing the charge per mass ratio to be consistent with the MHD-EPIC simulation and to better resolve the ion inertial length.


\section{Simulation results and comparison with observations}\label{section:res}
\subsection{Magnetopause crossing}
\citet{Kitamura:2016} calculated the LMN coordinates for the MMS3 magnetopause crossing. The L axis is [0.1974, 0.2013, 0.9594], the M axis is [0.1170, 0.9669, 0.2269], and the N axis is [0.9733, 0.1570, 0.1673] in the GSM coordinates. This LMN coordinate system is used in the present paper to compare simulation results with observations. 

To compare the simulation results with the MMS3 observations, we extract the simulation data from a virtual satellite, which has the same orbit and speed ($\sim$1.57~km/s) as MMS3. In the MHD-EPIC and Hall-MHD simulations, the ion-scale features (such as the current sheet thickness, the ion-scale flux ropes, and the reconnection ion diffusion region) are 16 times larger than in reality, and hence the virtual satellites in the simulations take 16 times longer time to fly across such features. To be consistent with the MHD-EPIC and Hall-MHD simulations, we also present the ideal-MHD simulation results in the same scales as the MHD-EPIC and Hall-MHD simulations. However, we note that there is not any physical reason behind the scaling of ideal-MHD simulation results. The ideal-MHD equations do not have any intrinsic scales, and the ion-scale structures in the ideal-MHD simulation only depend on the simulation grid resolution.

\subsubsection{Magnetopause location}
\begin{figure}
  \includegraphics[width=2\textwidth, trim=20cm 0cm 0cm 0cm]{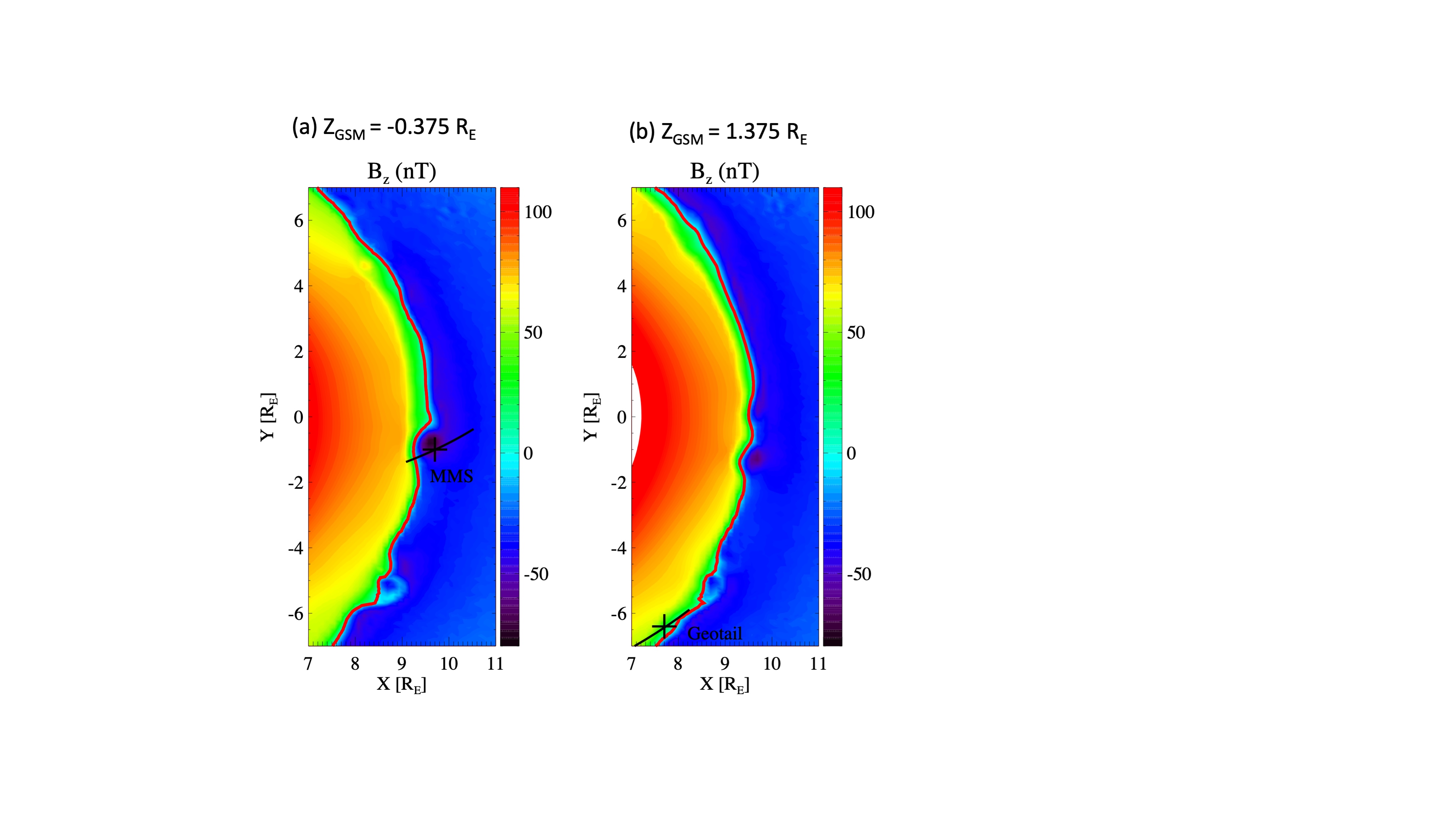}
  \caption{The $B_z$ magnetic field in the $Z_{GSM}=-0.375~R_E$ plane (left) and 
  the $Z_{GSM}=1.375~R_E$ plane (right) at the end of the MHD-EPIC simulation. 
  The magnetopause is identified by $B_{z,GSM} = 0$, which is the red line 
  in each of the plots. The MMS3 and Geotail satellites 
  were around [9.73, -0.98, -0.33] and [7.7, -6.4, 1.4] in GSM coordinates, respectively, when they
  acrossed the magnetopause. The black line and the black '+' sign in the left (right) figure represent 
  the MMS3 (Geotail) orbit and the observed magnetopause location that are projected onto the  
  $Z_{GSM}=-0.375~R_E$ ($Z_{GSM}=1.375~R_E$) plane. }
  \label{fig:mp}
\end{figure}

Figure~\ref{fig:mp} presents the $B_{z,GSM}$ magnetic field in the $Z_{GSM}=-0.375~R_E$ plane (left) and the $Z_{GSM}=1.375~R_E$ plane (right) at the end of the MHD-EPIC simulation. The red lines, where $B_{z,GSM}=0$, indicate the location of the simulation magnetopause. The black curves and the black '+' signs represent the satellite orbits and the observed magnetopause locations. The 'bumps' of the magnetopause (red lines) are produced by the reconnection effects. During the simulation, the magnetopause shape and location vary, but the distances between the satellites observed magnetopause locations (black '+') and the nearest simulation magnetopause are always within $0.5~R_E$, which can be verified by the the magnetopause crossing data in Figure~\ref{fig:plot1}. Figure~\ref{fig:plot1} plots the magnetic fields collected by the MMS3 satellite and the virtual satellites in the simulations. We note that the spatial and temporal scales of the simulation plots are 16 times larger than the MMS3 observations due to the scaling. In the MMS3 data, the magnetopause identified by $B_l=0$ is around $X_{GSM}=9.735~R_E$, and it is around $X_{GSM}=9.4~R_E$ for the MHD-EPIC simulation.

\subsubsection{Magnetic fields}
Figure~\ref{fig:plot1} shows the magnetopause crossing magnetic fields from the MMS3 spacecraft, the Auburn Hybrid model \citep{Guo:2020}, and the SWMF ideal-MHD, Hall-MHD and MHD-EPIC simulations. The Auburn hybrid model is another model that simulated the GEM dayside kinetic processes challenge event. We plot the hybrid simulation results here for completeness, and more details about the hybrid simulation can be found in \citet{Guo:2020}. We focus on the comparison between the MMS3 data and the SWMF simulations in the present paper. 

All the three SWMF simulations are essentially the same when the virtual satellites are far from the magnetopause. The magnitude of the magnetic field $B_t$ and the $B_l$ component from the SWMF simulations agree with MMS3 observations very well both in the magnetosphere (left end of Figure~\ref{fig:plot1}) and in the magnetosheath (right end of Figure~\ref{fig:plot1}). The $B_m$ component from the simulations also matches MMS3 data very well in the magnetosphere, but not in the magnetosheath. MMS3 observed a significant positive component of $B_m$ in the magnetosheath. However, the simulation $B_m$ is very close to zero in the magnetosheath, because the $B_m$ component is dominated by the $B_{y,GSM}$ component, and $B_{y,GSM}$ is zero in the simulation solar wind conditions. The difference in the $B_m$ component between the simulations and the MMS3 data may come from the simplified upstream IMF conditions. The $B_n$ component is essentially zero in both MMS3 observations and the simulations besides the small-scale oscillations. 

Across the current sheet (from $X_{GSM}=9.72~R_E$ to $X_{GSM}=9.74~R_E$ for MMS3), both the MMS3 and the MHD-EPIC $B_l$ components decrease at a similar rate from the magnetosphere to the magnetosheath. This suggests that suggests the MHD-EPIC simulation captures the current sheet thickness correctly. The Hall-MHD simulation shows a comparable decreasing rate, but it contains more large-amplitude oscillations than both the MMS3 data and the MHD-EPIC simulation. Since the current sheet structure of the ideal-MHD simulation strongly depends on the grid resolution, we will ignore the ideal-MHD simulation for the current sheet related comparisons. 

Around $X_{GSM}=9.72~R_E$, MMS3 observed a dip in $B_l$, $B_m$, and $B_t$, and the MHD-EPIC simulation also shows similar structures. A detailed comparison will be presented in section~\ref{section:fr}. Since the current sheet is quite dynamic, the simulations can not reproduce all features. For example, around $X_{GSM}=9.75~R_E$, MMS3 observed that the $B_l$ component field increases to zero, and the $B_m$ and $B_n$ components show significant variations, but none of the simulations capture these structures. 

\begin{figure}
  \includegraphics[width=1.2\textwidth, trim=0cm 0cm 0cm 0cm]{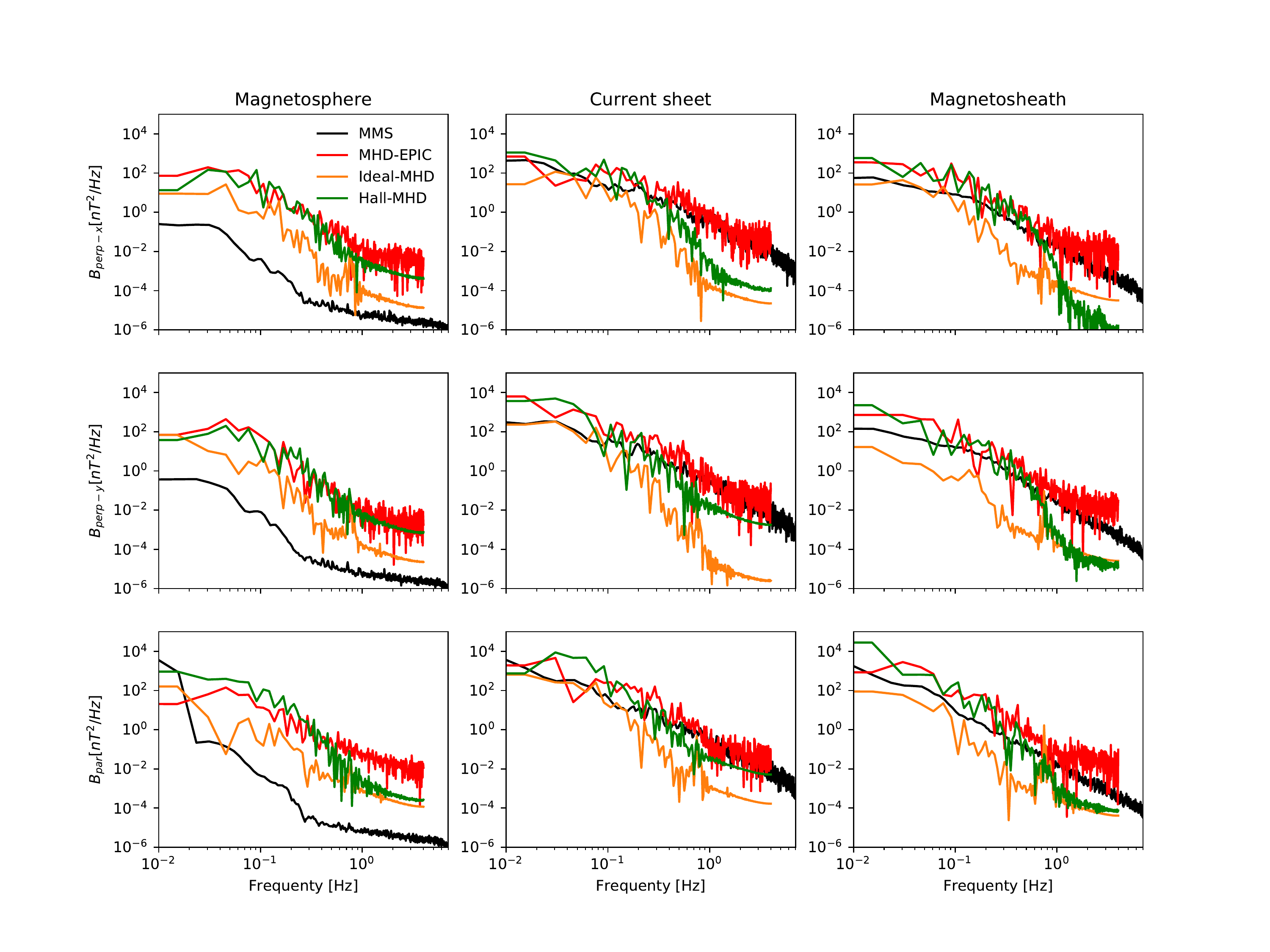}
  \caption{The power spectral densities (PSDs) of the parallel and perpendicular magnetic field components 
  in the magnetosphere (left column), the current-sheet (middle column), and the magnetosheath (right column). 
  The PSDs of the simulations are scaled by the scaling factor 16.}
  \label{fig:psd}
\end{figure}

Figure~\ref{fig:psd} shows the power spectral densities (PSDs) of the perpendicular and parallel magnetic field fluctuations in the magnetosphere, the current sheet, and the magnetosheath. The details of calculating the PSDs from the MMS3 data can be found in \citet{Guo:2020}. In the simulations, we use the magnetic field data collected at $X_{GSM}=8.82~R_E$, $X_{GSM}=9.34~R_E$ and $X_{GSM}=9.83~R_E$ along the MMS3 orbit to represent the magnetosphere, current sheet, and magnetosheath, respectively. $B_l$ is the parallel component, $B_m$ and $B_n$ are the two perpendicular components. Since the ion temporal scales in the MHD-EPIC and pure Hall-MHD simulations are 16 times slower than the reality due to the scaling, the simulation PSDs in Figure~\ref{fig:psd} are scaled by a factor of 16 to match the MMS3 data. The MHD-EPIC PSDs agree with observations well in the current sheet and the magnetosheath in general, but they are much higher than the MMS3 PSDs in the magnetosphere. Even though there is a significant difference in the magnetosphere PSDs between the MMS3 data and the MHD-EPIC simulation, both show the same trend that the magnetosphere PSDs are much smaller than either the current sheet or magnetosheath PSDs for the high frequencies ($>$ 0.1 Hz). The Hall-MHD PSDs are very similar to the MHD-EPIC PSDs for the frequencies less than 1~Hz, and they decrease faster than the MHD-EPIC PSDs for the frequencies larger than 1~Hz in general. The ideal-MHD PSDs are also presented for completeness, but we note again that the ideal-MHD PSDs strongly depend on the numerical parameters. 

\subsubsection{Ion profiles}
Figure~\ref{fig:plot2} shows the ion density, temperatures, and velocities during the magnetopause crossing. With an inner boundary density of 8~amu/cc, the ion densities of the SWMF simulations on the magnetospheric side match the MMS3 observation well. The simulation densities in the magnetosheath also agree with MMS3 data due to the proper simulation solar wind plasma density. The density variations around $X_{GSM} = 9.72~R_E$ are probably caused by flux rope-like structures. Section~\ref{section:fr} shows such structures in detail. 

The temperatures from all three SWMF simulations match MMS3 data in the magnetosheath. The MHD-EPIC parallel temperature also matches the observation very well in the magnetosphere, but the MHD-EPIC perpendicular temperature is just about 1400~eV while the observed value is about 2000~eV. The Hall-MHD and ideal-MHD magnetospheric temperatures are about twice higher than the MMS3 data. We note that the temperature is a scalar in the Hall-MHD and ideal-MHD simulations, and the parallel and perpendicular temperatures are the same.

MMS3 observed high-speed southward flow between $X_{GSM}=9.72~R_E$ and $X_{GSM}=9.74~R_E$. The flow reached a velocity of $v_{i,l} \approx -300~km/s$. This fast ion flow is likely to be the product of magnetic reconnection. The simulations also show such ion jets, but the simulation jets only reach a velocity of $v_{i,l} \approx -200~km/s$. The outflow velocity calculated from the Cassak-Shay equation \citep{Cassak:2007} is 190~km/s by choosing the magnetosheath and magnetosphere densities and magnetic fields $n_{i,sp}=1$~amu/cc, $n_{i,sh}=35$~amu/cc, $B_{t,sp}=60$~nT, and $B_{t,sh}=30$~nT, where the subscript 'sh' indicates the magnetosheath, and 'sp' represents the magnetosphere. The simulated outflow velocity is very close to the velocity from the Cassak-Shay equation. The MMS3 also observed jets between $X_{GSM}=9.74~R_E$ and $X_{GSM}=9.76~R_E$, but the simulations do not produce similar structures. The most significant difference between the observations and the simulations is the $v_{i,m}$ component in the magnetosphere. The MMS3 observed a velocity of $v_{i,m} \approx 250~km/s$, but none of the simulations produce such high velocity. Since the virtual satellites are around $Y_{GSM}\approx -1~R_E$, which is close to the meridian plane, during the magnetopause crossing, it is reasonable that the simulations do not produce large $v_{i,m}$ component. The difference between the simulations and the MMS3 data is unknown so far. 

\subsubsection{Electron profiles}
\begin{figure}
  \includegraphics[width=1.2\textwidth, trim=0cm 0cm 0cm 0cm]{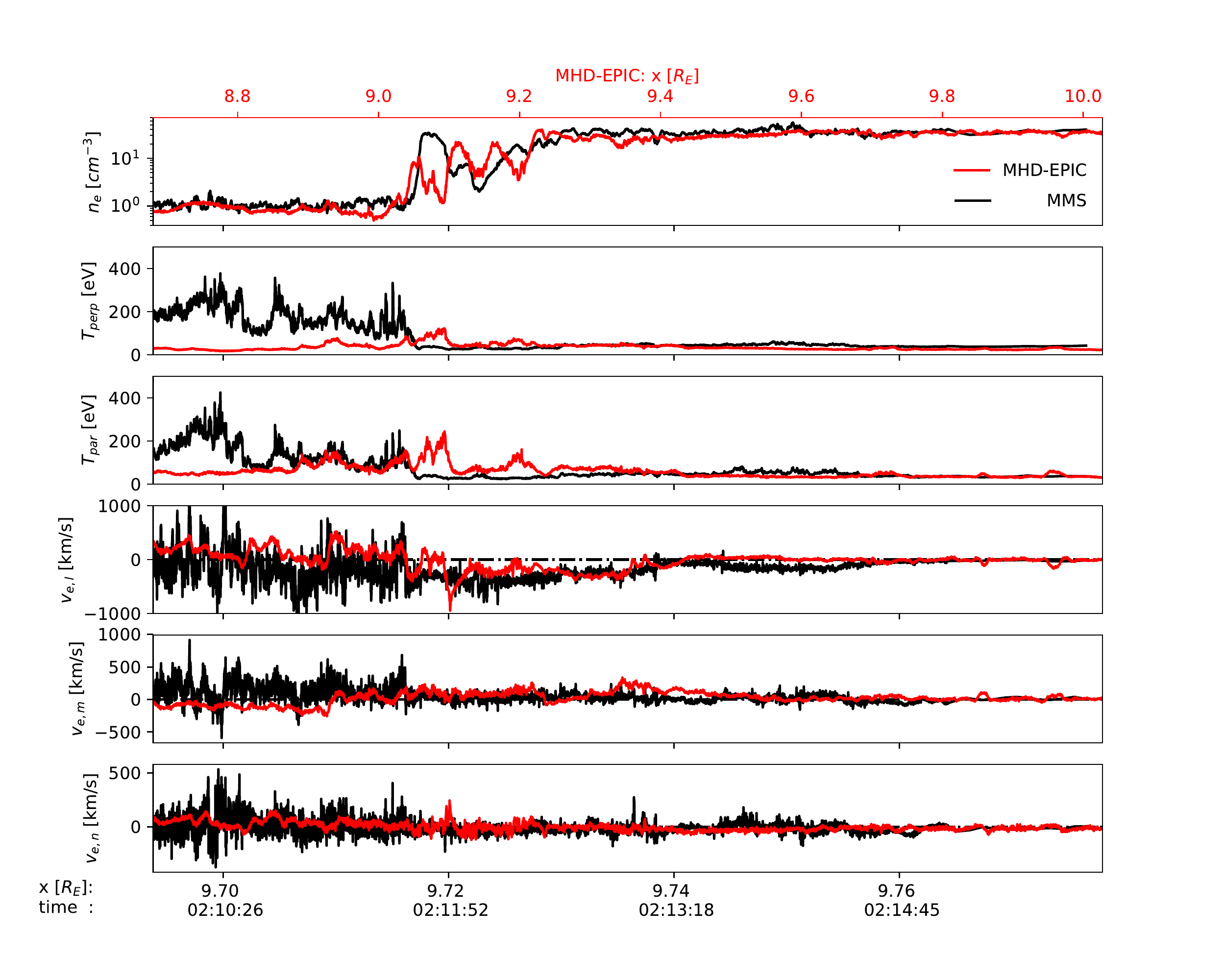}
  \caption{The electron profiles from the MMS3 spacecraft and the MHD-EPIC simulation. 
  The bottom X-axis indicates the $X_{GSM}$-coordinate and the time for the MMS3 observations, and the upper red X-axis 
  represents the $X_{GSM}$-coordinate for the MHD-EPIC simulation.}
  \label{fig:plot2-el}
\end{figure}

Since the MHD-EPIC model can provide electron information, Figure~\ref{fig:plot2-el} plots the electron data. The electron density is essentially the same as the ion density for both the MHD-EPIC simulation and the MMS3 observation due to charge neutrality at scales much larger than the Debye length. The MHD-EPIC electron temperatures agree with MMS3 data in the magnetosheath. But the simulated electron temperatures are lower than the observations in the magnetosphere, especially for the perpendicular temperature. In the electron velocity profiles observed by the MMS3 spacecraft, there are a lot of small-scale high-amplitude oscillations. Such oscillations are missing in the MHD-EPIC simulation probably due to the limitations of the grid resolution and time step. Between $X_{GSM}=9.72~R_E$ and $X_{GSM}=9.74~R_E$, the MMS3 spacecraft observed an electron jet velocity of $v_{e,l}\approx -500~km/s$. The MHD-EPIC simulation also produces electron jets with a similar velocity.

\begin{figure}
  \includegraphics[width=1.2\textwidth, trim=0cm 0cm 0cm 0cm]{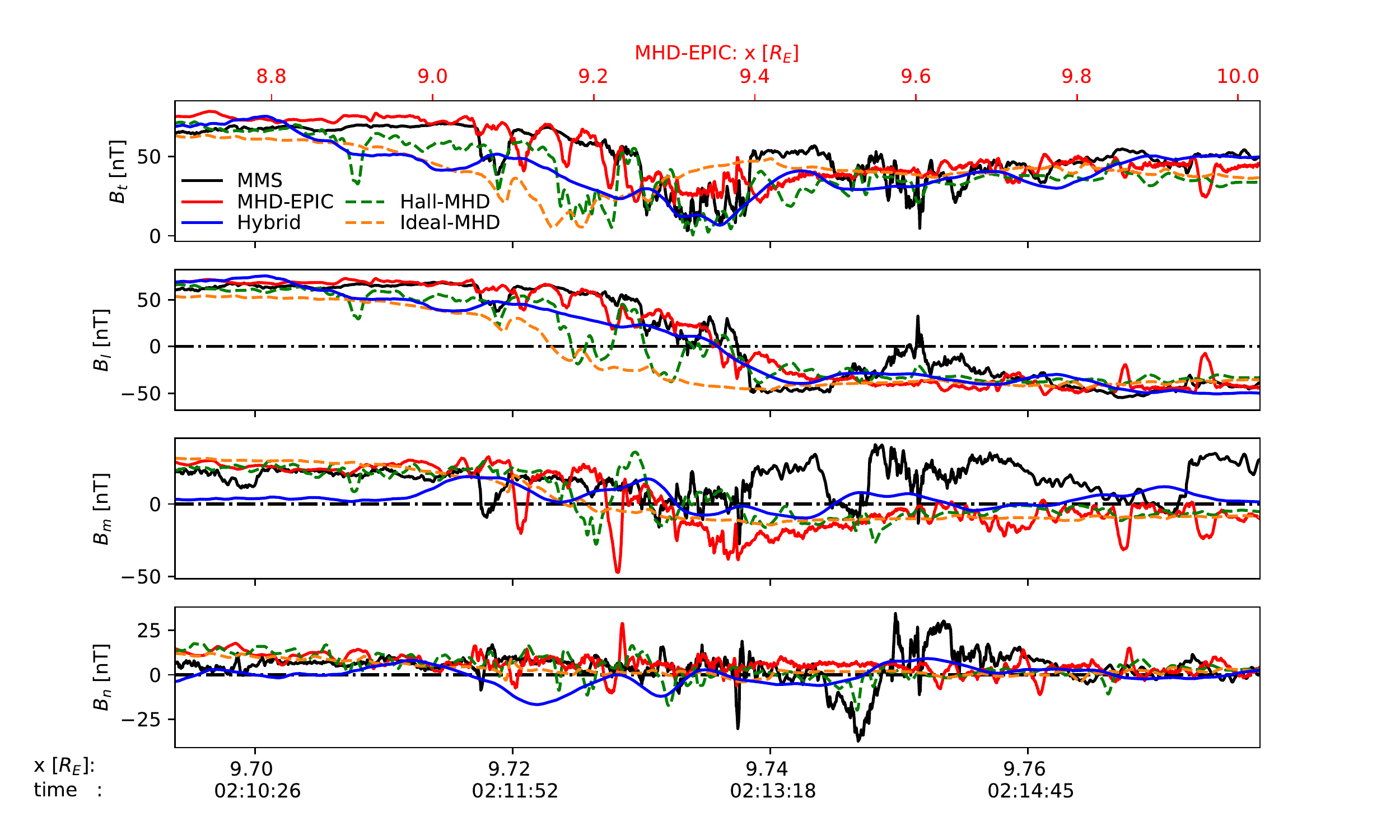}
  \caption{The magnetopause crossing magnetic fields from the MMS3 spacecraft, the Auburn hybrid model, 
  and the SWMF ideal-MHD, Hall-MHD and MHD-EPIC simulations.
  The MMS3 data from t=2:10:00 to t=2:16:00 is plotted. The bottom X-axis indicates 
  the $X_{GSM}$-coordinate and the time for the MMS3 observations and the Hybrid model.
  The upper red X-axis shows the $X_{GSM}$-coordinate for the ideal-MHD, Hall-MHD, and MHD-EPIC simulations. 
  The spatial and temporal scales of the SWMF simulations are 16 times larger than the MMS3 observations due 
  to the scaling. $B_t$ is the total field magnitude, while $B_l$, $B_m$ and $B_n$ are the 3 components in the LMN coordinate system.}
  \label{fig:plot1}
\end{figure}

\begin{figure}
  \includegraphics[width=1.2\textwidth, trim=0cm 0cm 0cm 0cm]{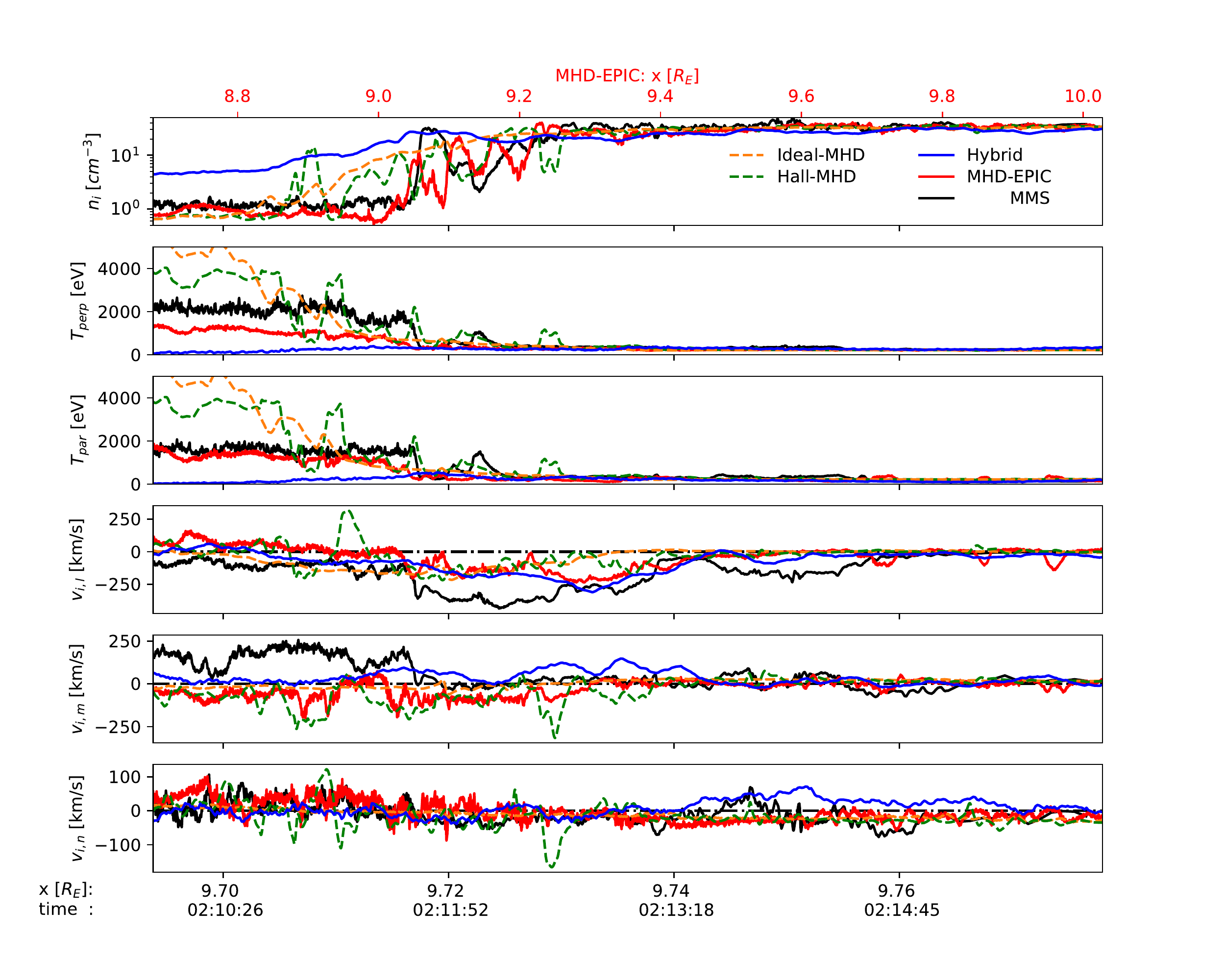}
  \caption{The ion profiles from the MMS3 spacecraft, the Auburn Hybrid model, and the SWMF ideal-MHD, Hall-MHD and 
  MHD-EPIC simulations. The X-axes are the same as those of Figure~\ref{fig:plot1}.}
  \label{fig:plot2}
\end{figure}

\subsubsection{Flux ropes during the magnetopause crossing}
\label{section:fr}
\begin{figure}
  \includegraphics[width=0.7\textwidth, trim=0cm 0cm 0cm 0cm]{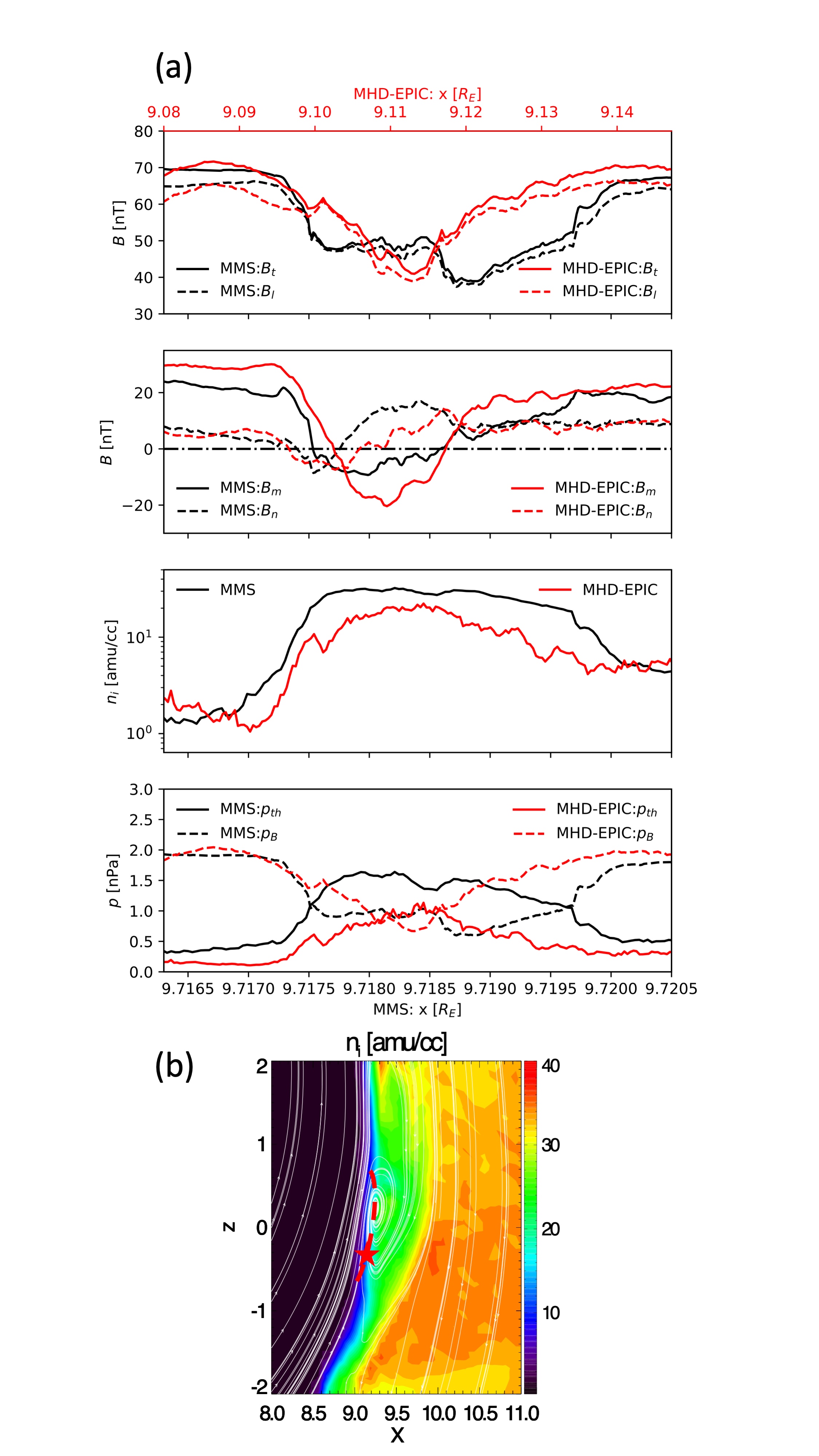}
  \caption{(a) The comparisons of the magnetic field, the ion density $n_i$, the plasma pressure ($p_{th}$) and 
  the magnetic field pressure ($p_B$) of an FTE from the MMS observations (black lines) and the MHD-EPIC simulation (red lines).
  The lower (upper) X-axis represents the coordinate for the MMS (MHD-EPIC) data. (b) The plasma density and magnetic field lines in the $Y_{GSM}=-1.437~R_E$ plane. The red star indicates the location of the virtual satellite when the virtual satellite is at  $X_{GSM}=9.1~R_E$. The red dashed line illustrates how the flux rope moves across the virtual satellite. We note that the red dashed line is not the virtual satellite orbit. }
  \label{fig:fr}
\end{figure}

The magnetic fields and density variations observed by the MMS3 spacecraft between $X_{GSM}=9.715~R_E$ and $X_{GSM}=9.72~R_E$ can match the signatures of a flux rope. Figure~\ref{fig:fr}(a) shows the magnetic fields and plasma profiles from both the MMS3 data and the MHD-EPIC simulation. Compared to Figure~\ref{fig:plot1} and Figure~\ref{fig:plot2}, the MHD-EPIC data in Figure~\ref{fig:fr}(a) is shifted a little bit in order to directly compare with MMS3 data. Figure~\ref{fig:fr}(b) illustrates how the corresponding flux rope moves across the virtual satellite in the MHD-EPIC simulation. When the virtual satellite is still in the magnetosphere, the bulge of a flux rope propagates through the virtual satellite. Since the virtual satellite is always on the magnetospheric edge of the flux rope, $B_l$ is always positive during the flux rope crossing, but the value of $B_l$ decreases when the virtual satellite moves closer to the flux rope center. The $B_n$ component changes sign even though the negative part of the $B_n$ field is not significant. The virtual satellite observes a core field of $B_m \approx -15$~nT near the center of the flux rope. The virtual satellite observes significant enhancements of plasma density and plasma thermal pressure inside the flux rope, since it moves from the magnetosphere into the magnetosheath. It is a southward propagating flux rope that produces all of the features in the simulation. Figure~\ref{fig:fr}(b) shows the corresponding flux rope. The MMS3 data presents similar structures, so it is likely the MMS3 spacecraft also observed a flux rope. 

\subsection{Movement and spreading of the X-lines}
 To compare the movement and spreading of the X-lines with observations, we design an automatic algorithm to identify X-lines based on the MHD-EPIC simulation electron jets velocities. First, we extract the 2D magnetopause surface from the PIC outputs by selecting the surface of $B_{z,PIC}=0$. Secondly, on the magnetopause surface, we loop through each column of the cells from the $-Z_{PIC}$ direction to the $+Z_{PIC}$ direction, and find out the location $Z^{'}_{PIC}$, where the electron velocity $v_{e,z}$ changes from southward (negative) to northward (positive). Finally, the velocity difference $\Delta v_{e,z}$ between the maximum and minimum electron velocity $v_{e,z}$ within $Z_{PIC} \in [Z^{'}_{PIC}-\Delta z, Z^{'}_{PIC}+\Delta z]$ is calculated. If $\Delta v_{e,z}$ is larger than the threshold value $\Delta v_{threshold}$, the location $Z^{'}_{PIC}$ is identified as a reconnection site. In this section, we choose $\Delta z = 0.4~R_E$, which is about 4 times of the magnetosheath ion inertial length, and $\Delta v_{threshold} = 200~km/s$, which is close to the magnetosheath Alfven speed. This simple algorithm is not very sensitive to the choices of $\Delta z$ and $\Delta v_{threshold}$. For example, changing the parameters to $\Delta z = 0.6~R_E$ and $\Delta v_{threshold} = 300~km/s$ will not alter the results too much. Since the PIC simulation coordinates are not parallel with the GSM coordinates, we present the PIC simulation results in its simulation coordinate system in this section.

\begin{figure}
  \includegraphics[width=0.95\textwidth, trim=0cm 0cm 0cm 0cm]{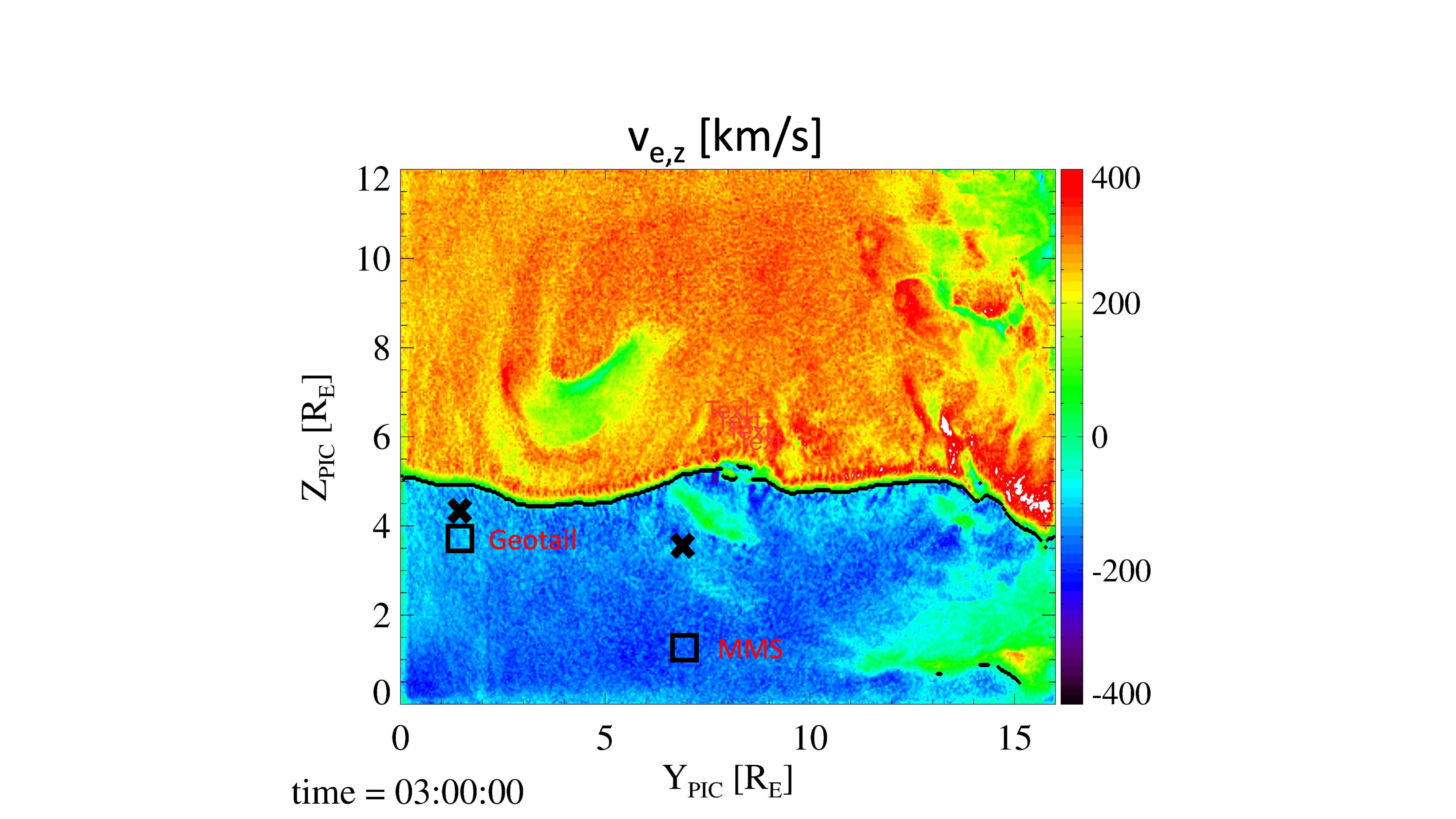}
  \caption{The electron velocity $v_{e,z}$ on the magnetopause in the PIC simulation coordinates at t=03:00:00. The black lines represent the simulation X-lines. The black squares represent the locations of the satellites when they observed the magnetopause, and the black crosses indicate the X-line locations that are estimated from the satellite data \citep{Kitamura:2016}. }
  \label{fig:xline1}
\end{figure}

\begin{figure}
  \includegraphics[width=2.5\textwidth, trim=0cm 0cm 0cm 0cm]{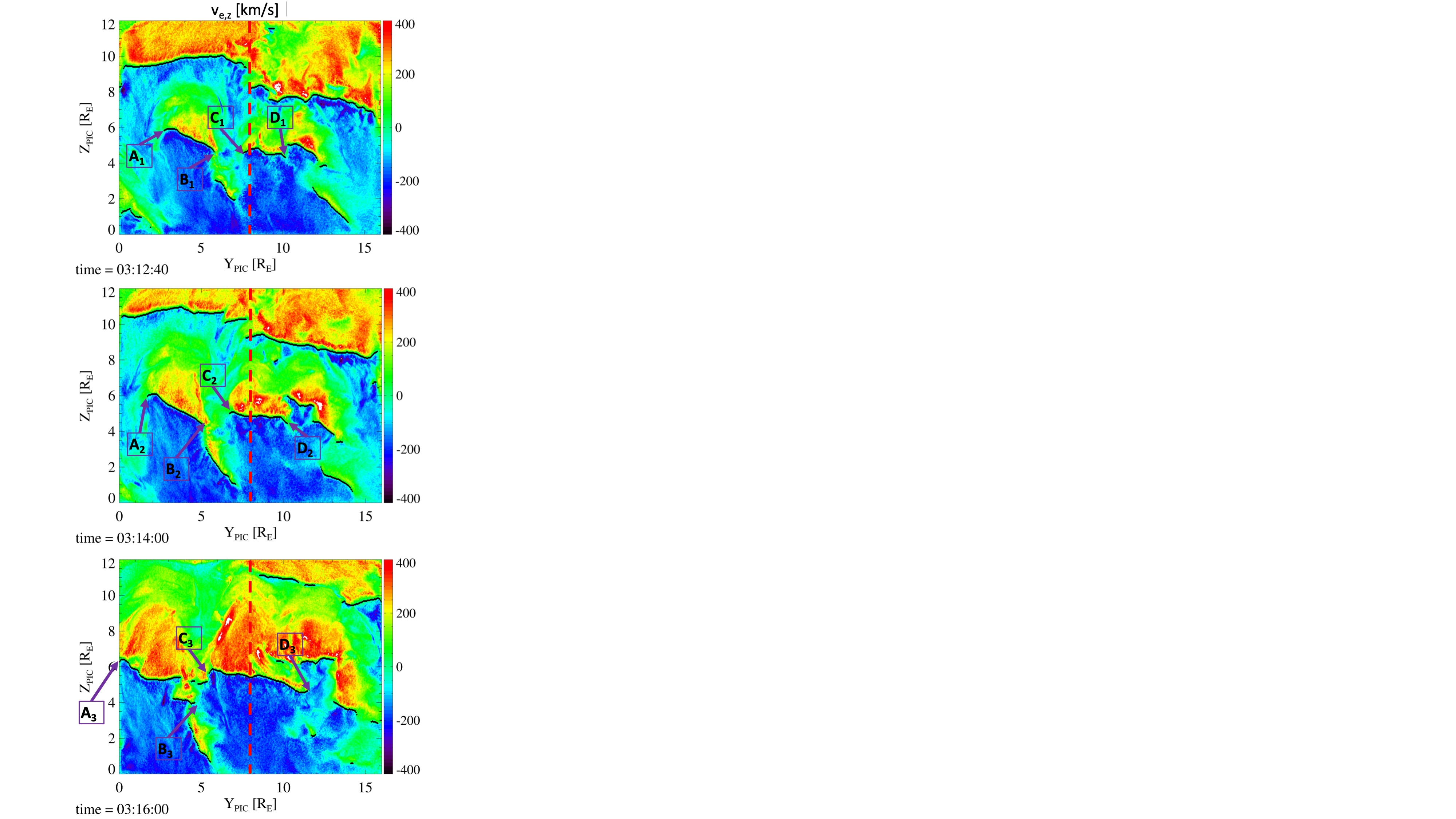}
  \caption{The evolution of the X-lines on the magnetopause. The vertical red dashed lines indicate the location of noon.}
  \label{fig:xline2}
\end{figure}

An example of the X-lines identified by the algorithm is presented in Figure~\ref{fig:xline1}. There is a long X-line at this moment. This X-line is around $Z_{GSM} \approx 3~R_E$ in the GSM coordinates due to the tilting of the dipole field, which is consistent with the MMS3 and Geotail observations by \citet{Kitamura:2016}. However, it is unusual to form such a long single X-line in the MHD-EPIC simulation. It is more typical to have multiple X-lines at the same time in the PIC simulation domain, just as what is shown in Figure~\ref{fig:xline2}. 

In the MHD-EPIC simulation, the evolution of the X-lines, which are identified by the algorithm described above, is very dynamic and complicated. We will systematically analyze the evolution of the X-lines in detail in a forthcoming paper. The following part of this section presents some examples that may be related to the X-line spreading observed by \citet{Zou:2018}. 

By tracing the locations of the X-line edges, we can study the movement and spreading of the X-lines. Points A, B, C and D in Figure~\ref{fig:xline2} indicate the ends of two X-lines. Table~\ref{tb} shows the locations and moving speeds of the end points at $t_1$=03:12:40, $t_2$=03:14:00, and $t_3$=03:16:00. The subscripts of points A, B, C and D indicate the time. The speeds are estimated based on the motion between two snapshots. Points A and B are the left and right edges of an X-line, respectively. Point A moves dawnward with a speed of $\sim 80$~km/s, and Point B also moves dawnward but with a slightly slower speed of $\sim 64$~km/s. Since the speed difference between points A and B is very small, the X-line between A and B moves dawnward and its length does not grow too much. At $t_3$, the X-line between A and B has already split into two X-lines. The X-line between points C and D is another example to show the growth of the X-line. From $t_1$ to $t_2$, point C moves dawnward at a speed of $\sim 60$~km/s, and point D does not move too much. So, this X-line spreads dawnward between these two snapshots. From $t_2$ to $t_3$, point D also moves duskward fast with a speed of  $\sim 70$~km/s, and this X-line spreads at both ends. The length of the X-line between points C and D grows from $2.5~R_E$ at $t_1$ to $6~R_E$ at $t_3$. These examples suggest that the typical propagation speed of an X-line endpoint is about 70~km/s. If both endpoints of an X-line move towards the same direction at the same speed, it behaves like the whole X-line moves in one direction. If one X-line endpoint is steady or the two endpoints move in the opposite directions, the X-lines spreads in one direction or both directions. 

\citet{Zou:2018} found that the total spreading speed of the X-lines under a weak guide field is about 40~km/s. Even though the spreading speeds obtained from the MHD-EPIC simulation are about 2 to 4 times faster than the observations, they are still comparable. The evolution of the X-lines can be very complicated, and we will present a systematic investigation in the forthcoming paper. 

\begin{table}[ht]
  \label{tb}
  \caption{The locations and speeds of the X-line endpoints that are marked in Figure~\ref{fig:xline2}.
  $t_1$=03:12:40, $t_2$=03:14:00, and $t_3$=03:16:00. Speeds $v_{1,2}$ and $v_{2,3}$ are calculated from 
  the motion of the points from $t_1$ to $t_2$ and $t_2$ to $t_3$, respectively.}
  \centering
  \begin{tabular}{c c c c c c}
  \hline
   Point    & $Y_{PIC}$ at $t_1$ & $Y_{PIC}$ at $t_2$ & $Y_{PIC}$ at $t_3$ & $v_{1,2}$ [km/s] & $v_{2,3}$ [km/s]\\
  \hline
 A        & 2.8     & 1.8                 & 0     & 80  & 96  \\
 B        & 5.8     & 5                   & 3.8   & 64  & 64 \\
 C        & 7.5     & 6.8                 & 5.5   & 56  & 70  \\
 D        & 10      & 10.2                & 11.5  & 10 & 70 \\
 \hline
  \label{tb:parameters}
  \end{tabular}
  \end{table}

\section{Summary}
The MHD-EPIC model is used to study the southward IMF event on 2015-11-18 01:50-03:00 UT. The simulation results are compared with the satellite data and the ground-based SuperDARN observations. The key results are:
\begin{itemize}
  \item The magnetopause location obtained from the MHD-EPIC simulation is very close to the magnetopause location identified by either MMS3 or Geotail. Along the MMS3 orbit, the magnetopause observed by MMS3 is around $X_{GSM}=9.735~R_E$, and it is around $X_{GSM}=9.4~R_E$ in the MHD-EPIC simulation.
  \item The simulation magnetic fields match the MMS3 data very well except for the magnetosheath $B_m$ component. The discrepancy may be caused by the difference between the simulation IMF and the actual IMF. 
  \item The simulation ion density, perpendicular temperature, and parallel temperature match the MMS3 data well. Both the simulation and the MMS3 spacecraft observed southward high-speed ion flow. 
  \item The MHD-EPIC simulation provides electron information. The simulation electron number density agrees with MMS3 data well, but the simulation temperatures in the magnetosphere are lower than the MMS3 data. Both the MMS3 data and the simulation present electron jets with a velocity of $v_{e,l}\approx -500$~km/s.
  \item  The MHD-EPIC simulation produces FTEs. The magnetic field and plasma variations between $X_{GSM}=9.716~R_E$ and $X_{GSM}=9.72~R_E$ in the MMS3 data match the signatures of an FTE crossing event. 
  \item There are usually multiple X-lines in the simulation domain instead of one long X-line. 
  \item The movement and spreading of X-lines are identified from the MHD-EPIC simulation. The endpoints of an X-line usually move at a speed of $\sim70$~km/s, which is about 2 to 4 times faster than the SuperDARN observed X-line spreading speed.
\end{itemize} 
Overall the MHD-EPIC simulation results show good agreement with observations, and in general this model agrees better than the simpler Hall MHD and ideal MHD models. The results suggest that MHD-EPIC can reproduce both the global and the small scale structures successfully.

\acknowledgments
This work was supported by the INSPIRE NSF grant PHY-1513379 and the NSF
PREEVENTS grant 1663800. Computational resources supporting this work were
provided on the Frontera super computer through the Texas Advanced Computing Center,
on the Pleiades computer by NASA High-End Computing (HEC) Program through
the NASA Advanced Supercomputing (NAS) Division at Ames Research Center,
and from Cheyenne (doi:10.5065/D6RX99HX) provided by NCAR's Computational and
Information Systems Laboratory, sponsored by the National Science Foundation.

The MMS datasets are publicly available at the MMS Science Data 
Center at https://lasp.colorado.edu/mms/sdc/public/.
The SWMF code (including BATS-R-US and iPIC3D) is publicly available through the
csem.engin.umich.edu/tools/swmf web site after registration.
The simulation output used for generating the figures in this paper can be obtained
via https://umich.box.com/s/g74ild6z4wd4klcqv4u84cge0gkkncwf. 

\clearpage
\bibliography{csem,yuxichen}

\end{document}